\documentclass[aps,prl,reprint,showpacs,superscriptaddress,longbibliography]{revtex4-1}
\usepackage{mathtools,amssymb,graphicx,units}

\usepackage[plainpages=false,pdfpagelabels,colorlinks=true,linkcolor=red,urlcolor=blue,citecolor=blue,pdftitle={Title},pdfauthor={},pdfdisplaydoctitle=true,pdfduplex=DuplexFlipLongEdge]{hyperref}
\usepackage{verbatim}
\usepackage[english]{babel}  

\begin{document}

\title{Observation of energy resolved many-body localization}

\author{Qiujiang Guo}
\thanks{Those authors contributed equally to this work.}
\affiliation{Interdisciplinary Centre for Quantum Information and Zhejiang Province Key
	Laboratory of Quantum Technology and Device, Department of Physics, Zhejiang University,
	Hangzhou 310027, China}

\author{Chen Cheng}
\thanks{Those authors contributed equally to this work.}
\affiliation{School of Physical Science and Technology, Lanzhou University, Lanzhou 730000, China}
\affiliation{Beijing Computational Science Research Center, Beijing 100094, China}

\author{Zheng-Hang Sun}
\thanks{Those authors contributed equally to this work.}
\affiliation{Institute of Physics, Chinese Academy of Sciences, Beijing 100190, China}
\affiliation{CAS Center for Excellence in Topological Quantum Computation, School of Physical Sciences, UCAS, Beijing 100190, China}

\author{Zixuan Song}
\author{Hekang Li}
\author{Zhen Wang}
\author{Wenhui Ren}
\author{Hang Dong}
\affiliation{Interdisciplinary Centre for Quantum Information and Zhejiang Province Key
	Laboratory of Quantum Technology and Device, Department of Physics, Zhejiang University,
	Hangzhou 310027, China}

\author{Dongning Zheng}
\affiliation{Institute of Physics, Chinese Academy of Sciences, Beijing 100190, China}
\affiliation{CAS Center for Excellence in Topological Quantum Computation, School of Physical Sciences, UCAS, Beijing 100190, China}

\author{Yu-Ran Zhang}
\affiliation{Theoretical Quantum Physics Laboratory, RIKEN Cluster for Pioneering Research, Wako-shi, Saitama 351-0198, Japan}

\author{Rubem Mondaini}
\email{rmondaini@csrc.ac.cn}
\affiliation{Beijing Computational Science Research Center, Beijing 100094, China}

\author{Heng Fan}
\email{hfan@iphy.ac.cn}
\affiliation{Institute of Physics, Chinese Academy of Sciences, Beijing 100190, China}
\affiliation{CAS Center for Excellence in Topological Quantum Computation, School of Physical Sciences, UCAS, Beijing 100190, China}

\author{H. Wang}
\email{hhwang@zju.edu.cn}
\affiliation{Interdisciplinary Centre for Quantum Information and Zhejiang Province Key
Laboratory of Quantum Technology and Device, Department of Physics, Zhejiang University,
Hangzhou 310027, China}

\begin{abstract}
{\bf Many-body localization (MBL) describes a quantum phase where an isolated interacting system subject to sufficient disorder displays non-ergodic behavior, evading thermal equilibrium that occurs under its own dynamics. Previously, the thermalization-MBL transition has been largely characterized with the growth of disorder. Here, we explore a new axis, reporting on an energy resolved MBL transition using a 19-qubit programmable superconducting processor, which enables precise control and flexibility of both disorder strength and initial state preparations. We observe that the onset of localization occurs at different disorder strengths, with distinguishable energy scales, by measuring time-evolved observables and many-body wavefunctions related quantities. Our results open avenues for the experimental exploration of many-body mobility edges in MBL systems, whose existence is widely debated due to system size finiteness, and where exact simulations in classical computers become unfeasible.
}
\end{abstract}

\maketitle


The phenomenon of MBL represents one of the paradigmatic examples of a typical out-of-equilibrium quantum phase transition \cite{Nandkishore2015,Altman2018,Abanin2019,Alet2018}. It goes beyond the standard ground-state classification of the quantum matter, and its associated low-lying excitations. Instead, it is described by a high-energy phase transition, inherently manifested via the unitary dynamics of an isolated system, wherein by tuning the strength of disorder, one is able to halt the onset of ergodic behavior and thermalization \cite{Deutsch1991,Srednicki1994,Rigol2008,Alessio2016}. As a direct consequence, the ability MBL systems possess in retaining information, naturally paves the way for a quantum information storage device.
A fundamental ingredient of the MBL is the interplay of disorder and interactions, in contrast with a typical non-interacting Anderson localization \cite{Anderson1958}, and hence the intricate balance of these two knobs triggers the onset of the out-of-equilibrium quantum phase transition.

\begin{figure*}[h!]
 \includegraphics[width=0.85\textwidth]{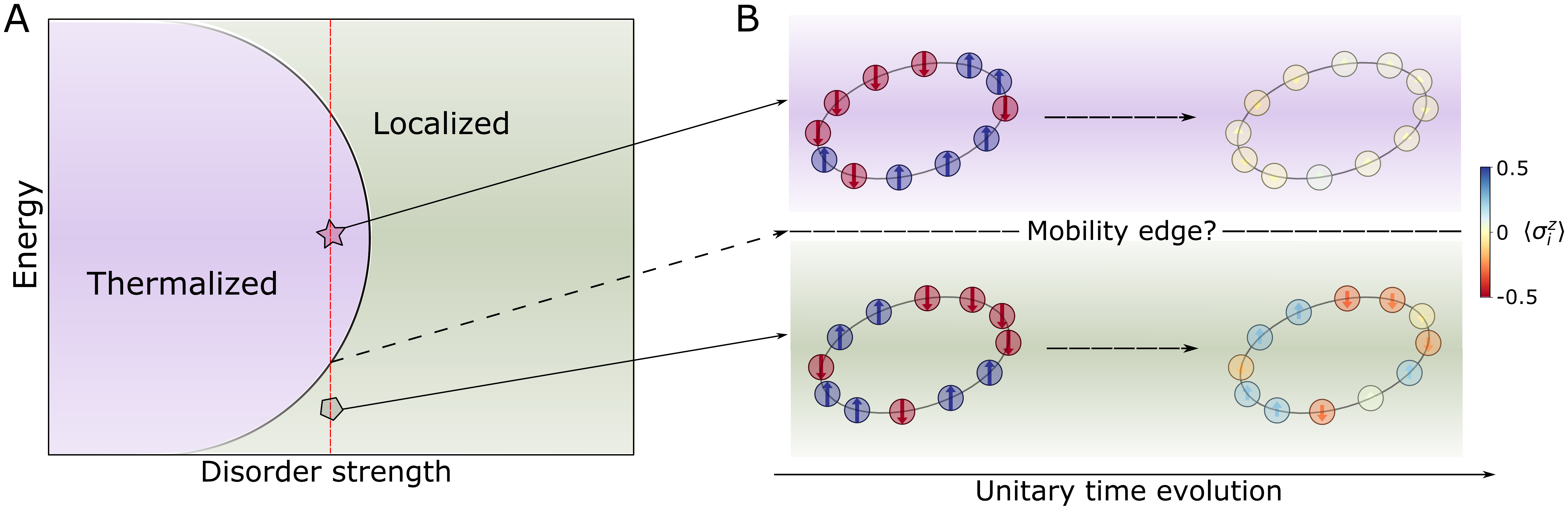}
 \vspace{-0.2cm}
\caption{{\bf Energy-resolved many-body localization and its measuring scheme.} {\bf (A)} Schematic nonequilibrium phase diagram in the energy-disorder space for an isolated interacting system. Despite of the ongoing debate, a finite system can be either in the thermalized or in the localized phase at the same disorder strength. {\bf (B)} Experimental scheme of measuring energy dependent localization by examining unitary time-evolution from different initial states.  A many-body mobility edge is suggested if contrasting results can be observed at long times for the quench dynamics with the {\it same} disorder strength but different energies. A colored ball with arrow denotes a two-level site, for example, a local spin; both color depth and arrow length describe the onsite magnetization.}
\label{fig:Fig0}
\end{figure*}

The recent high-precision experimental simulation of quantum many-body systems, engineered to operate quantum mechanically up to large time scales compared to the characteristic equilibration times, have demonstrated the capability of observation of this phenomenon, originally described by analytical \cite{Basko2006,Imbrie2016}, and numerical means \cite{Znidaric2008,Luitz2015,Mondaini2015,Wahl2019}. As a typical dynamical phase transition, the experimental investigations largely rely on the preparation of low entropy initial states -- often product states with high fidelity -- whose time-evolution is followed after a quench protocol. This was used to investigate the MBL transition, set off by growing disorder amplitudes, in experiments using ultracold atoms trapped by quasi-disordered optical lattices \cite{Schreiber2015,Bordia2016,Luschen2017,Kohlert2019,Rispoli2019}, or in disordered settings \cite{Choi2016}, as in the case of ion chains \cite{Smith2016} and, more recently, on quantum processors, emulated via transmon superconducting qubits \cite{Xu2018,Roushan2017,Neill2018}. In common, they follow a similar protocol: by checking how a few-body observable, tailored to be easily quantified on the initial state, evolves in time, a distinction can be drawn
between ergodic and localized behavior, naturally occuring below and above the critical disorder strength. However, experimental explorations of this transition have often neglected the energy dependence on the onset of localization, and the precise understanding of how the triggering of the MBL phase is influenced by this extra variable has direct consequences to potential technological applications on a quantum memory instrument. Figure~\ref{fig:Fig0} direct exemplifies a situation where information of the initial conditions can be or not preserved depending on the energy of the isolated system, evolved under unitary dynamics.

\begin{figure*}[h!]
 \includegraphics[width=0.95\textwidth]{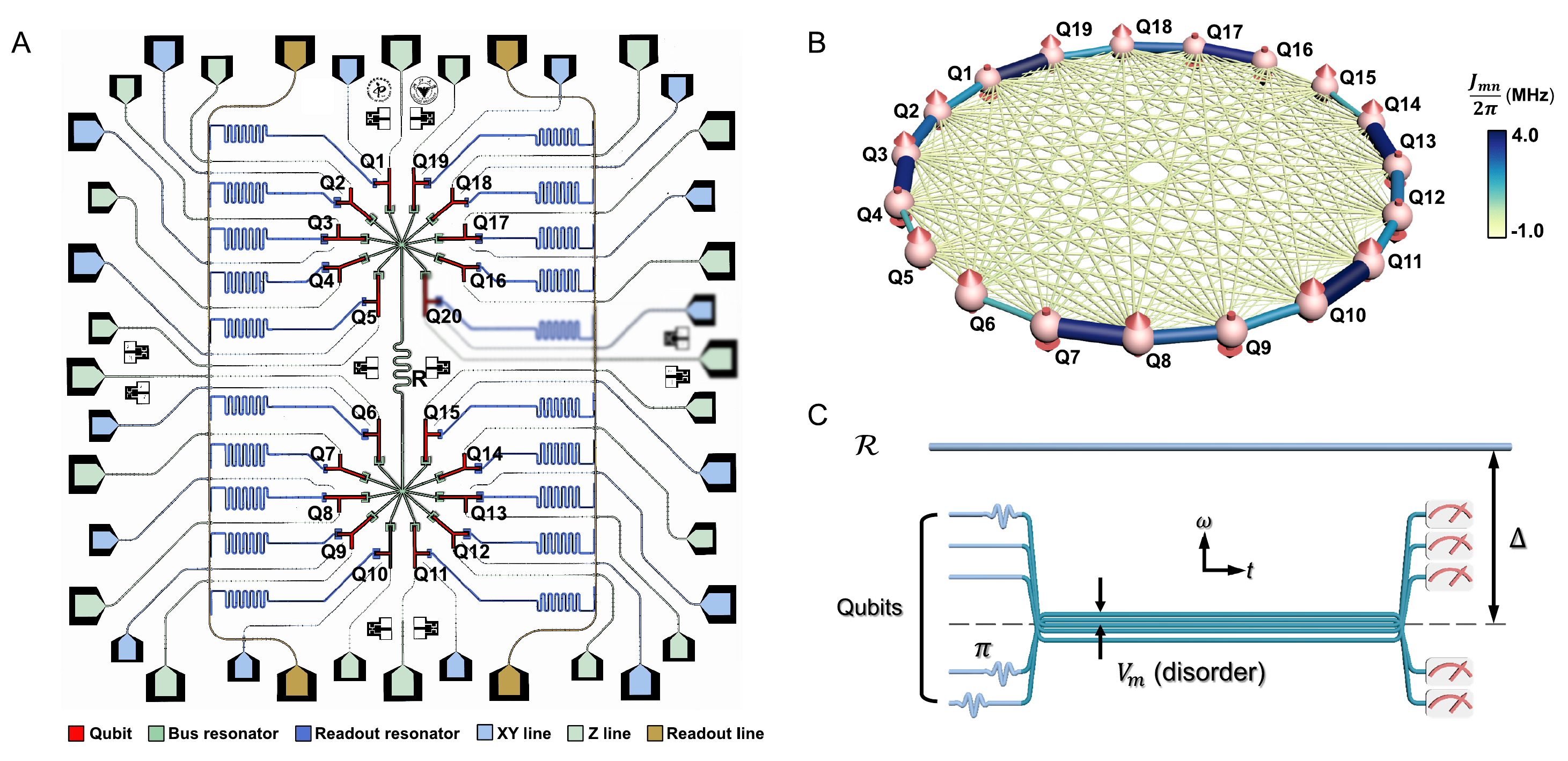}
 \vspace{-0.2cm}
 \caption{{\bf Quantum processor and experimental pulse sequence. (A)} False color image showing 20 superconducting qubits, labeled from $Q_1$ to $Q_{20}$, interconnected by the central bus resonator, {$\cal R$}. Critical circuit elements are rendered in different colors as indexed by the legend at bottom, where XY lines (light blue) are to excite qubits and Z lines (light green) are to tune qubit frequencies. Since $Q_{20}$'s frequency tunable range is relatively small, in this experiment we use 19 qubits by turning off $Q_{20}$'s coupling lines from the rest qubits. {\bf (B)} Schematic representation of the effective all-to-all coupling strengths, $J_{mn}$, among the 19 qubits (spins) while all qubits are placed at $\sim$568~MHz below the resonant frequency of {$\cal R$}. Both thickness and color depth of the connecting lines denote the magnitude of $J_{mn}$, which is applicable to this experiment (see Supplementary Material for more details).
{\bf (C)} Experimental pulse sequence in the frequency vs. time domain for observing energy resolved MBL. 19 qubits are initialized in $|0\rangle$ at their respective idle frequencies and specifically selected 9 qubits are then excited to $|1\rangle$ by $\pi$ pulses (sinusoids), following which all qubits are brought to around $|\Delta|$ below the resonant frequency of $\cal{R}$ for collective coherent dynamics. During the dynamics, each qubit $Q_m$ is slightly detuned from $\Delta$ by a small value of $V_m$, and to mimic a fully disordered system, we choose $V_m$ from a uniform random distribution [-$V$, $V$]. Combinations of the random disorder \{$V_m$\} and the correspondingly chosen 9 qubits excited at the beginning of the pulse sequence target a pre-set total energy $E$ (or the energy density $\varepsilon$) of the system. After the time interval $t$, all qubits are tuned to their respective readout frequencies for simultaneous multiqubit state measurement.}
\label{fig:Fig1}
\end{figure*}

Here, by using a newly designed 20-qubit quantum processor and flexibly programming $N = 19$ of the qubits (see Fig.~\ref{fig:Fig1}A), we are able to have a large control of interactions and disorder, which, combined with a precise initial state preparation, enables a direct probe of the energy-disorder phase diagram of the many-body localized phase transition with remarkable accuracy. Our insight is to perceive that in a quench problem, the initial state fully encodes the total energy of the system (within the regime the system is yet isolated from perturbations from the environment), and, as such, would allow to probe when localization occurs with energy resolution. That is, by initializing a product (Fock) state $|\Psi_0\rangle$, the unitary time-evolved wave function $|\Psi_t\rangle = e^{-{\rm i}Ht}|\Psi_0\rangle$ preserves the total energy  $\langle\Psi_t|H|\Psi_t\rangle=\langle\Psi_0|H|\Psi_0\rangle = E$, under the effective Hamiltonian of the superconducting quantum processor (see Supplementary Material)~\cite{Song2017,Xu2018,Song2019},
\begin{eqnarray}
\label{Ham}
\frac{H}{\hbar} &=& \sum_{\{m,n\}\in N} J_{mn}\left({\sigma^+_{m} \sigma_{n}^- + \sigma^-_{m} \sigma_{n}^+}\right) \nonumber \\
 && + \sum_{m} V_m \sigma^+_{m} \sigma_{m}^-,
\end{eqnarray}
where $\sigma^-_{m}$ ($\sigma^+_{m}$) is the lowering (raising) operator for qubit $Q_m$, and the first term runs at pairs of qubits $Q_m$ and $Q_n$.
The effective coupling strength $J_{mn}$ between pairs of qubits is schematically represented in Fig.~\ref{fig:Fig1}B, in which ``nearest-neighbor'' couplings are much larger. But the system possesses smaller long range couplings, resulting in a non-integrable Hamiltonian (Eq. 1) even in the absence of disorder \cite{Xu2018}. The second term in Eq. 1 is the disordered potential $V_m$ of the $m$-th qubit, which can be flexibly set by individually adjusting its resonant frequency, without noticeably altering the values of $J_{mn}$. Finally, to mimic a fully disordered system, we choose $V_m$ from a uniform random distribution $[-V,V]$.

The protocol we follow is: (i) With the $N=19$ qubits in $|0\rangle$, we prepare initial product states via $\pi$ pulses on individually selected $N_1 = 9$ qubits, producing generic states $|\Psi_0\rangle$. (ii) For a given disorder realization $\{V_m\}$, we compute the total energy of the system, where only diagonal terms in $H$ have a finite contribution: $E/\hbar=\sum\limits_{m \in N_1} V_{m}$, which sums over the 9 qubits initialized in $|1\rangle$. (iii) We estimate, for this $\{V_m\}$ realization, where the total energy $E$ lies in the energy density spectrum,
\begin{equation}
 \varepsilon = \frac{E-E_{\rm min}}{E_{\rm max}-E_{\rm min}}.
\end{equation}
The extremal eigenvalues of $H$, $E_{\rm min}$ and $E_{\rm max}$, are easilly obtainable by numerical means, without resorting to full diagonalization of $H$. (iv) As a direct quantification of the preservation of the information encoded in the initial state, we measure a few-body (and local) observable, the generalized imbalance described as
\begin{equation}
{\cal I}_{\rm gen} = \sum\limits_{m=1}^{19} \beta_m\sigma^+_m \sigma^-_m,
\end{equation}
where $\beta_m = 1/N_{1} \  (-1/N_{0})$ on the qubit $m$ initialized to $|1\rangle$ ($|0\rangle$); $N_1 =9$ and $N_0 = 10$ define the occupancy of our many-body system, where we select to excite 9 out of the $N=19$ qubits, in order to emulate the largest possible Hilbert space with our quantum processor, for a conserved total magnetization. The measurements are carried out for $k=20$ disorder realizations, together with carefully selecting the initial states, to fill a mesh of energy densities $\delta\varepsilon = 0.05$. This observable is similar to the charge imbalance often used in optical lattice experiments, when preparing a charge density wave \cite{Schreiber2015,Bordia2016,Luschen2017,Kohlert2019}, or a N\'eel state for trapped ions \cite{Smith2016}, with the advantage of being specifically customized for whichever initial state with $\langle\Psi_0|{\cal I}_{\rm gen}|\Psi_0\rangle=1$.

The experimental pulse sequence to execute the abovementioned protocol is shown in Fig.~\ref{fig:Fig1}C, which consists of $\pi$ pulses for exciting qubits, square pulses for tuning qubit frequencies, and simultaneous readout for obtaining multiqubit probabilities. Our single-qubit $\pi$ pulses are calibrated to be around 0.996 in fidelity using randomized benchmarking, which ensures that the state preparation is accurate. The readout fidelity values are around 0.97 (0.92) on average for $|0\rangle$ ($|1\rangle$) as measured simultaneously for all 19 qubits, which are used to correct the multiqubit probabilities for elimination of the readout errors before further processing of the data. Additionally, we can reliably obtain the quantum unitary evolution up to times $t =1500$~ns, which are much shorter than the typical energy relaxation times of the qubits (in the range from 30 to 70~$\mu$s) and much longer than the average shortest tunneling times $\tau = 1/({J_{m,m+1}^{\rm ave}}) \approx 60$~ns, sufficient to observe equilibration before decoherence processes ultimately affect the results (See Supplementary Material).

Figure \ref{fig:Fig2}{A} displays the disorder averaged generalized imbalance ${\cal I}_{\rm gen}$ at $t=1000$~ns: A characteristic `$D$'-shape structure on the energy density-disorder amplitude ($\varepsilon$-$V$) diagram highlights that loss of memory of the initial product states is both disorder- and energy-dependent in our $19$-qubit quantum processor, for a Hilbert space size with 92,378 states. Closely matching results can be obtained from exact numerical simulation using experimentally measured qubit couplings, and render a very similar diagram for ${\cal I}_{\rm gen}$ at equivalent time scales.

From numerical simulation, many different quantities can characterize the onset of the MBL transition \cite{Luitz2015,Mondaini2015}. The simplest is to look at the statistics of the gaps in the eigenenergy spectrum of $H$. Thermalization is intimately connected with the manifestation of quantum chaotic behavior, associated with eigenenergy level repulsion, whereas for MBL, disorder renders uncorrelated energies, resulting in Poissonian distribution of level spacings \cite{Alessio2016}. Experimentally, a spectroscopic analysis is rather elusive due to the exponentially dense energy spectra, but it was done on a 9-qubit device, when dealing with up to 45 states forming the Hilbert space \cite{Roushan2017}. One commonly tracks the average value of the ratio of adjacent gaps $r_\alpha=\min(\delta_\alpha,\delta_{\alpha+1})/\max(\delta_\alpha,\delta_{\alpha+1})$, with $\delta_\alpha$ a gap in the energy spectrum between consecutive levels $E_\alpha$ and $E_{\alpha+1}$; ergodic and localized regimes possess average values $\langle r\rangle$  approximately equal to 0.53 and 0.39, respectively \cite{Atas2013}. By restricting this analysis to different regions of the spectrum, we can construct the simulated MBL phase diagram for our case (Fig.~\ref{fig:Fig2}B), after using the experimentally measured $J_{mn}$. The striking similarity between the experimental and simulated phase diagrams, reassures the indication of energy-dependent localization transition. Focusing on the generalized imbalance at intermediate disorder ($V/2\pi= 16~{\rm MHz}$) in Fig.~\ref{fig:Fig2}C,  experimental and numerical results exhibit a remarkable agreement, enabling us to infer that our finite system prepared at low energy densities ($\varepsilon = 0.15$ and $0.3$) will fail to thermalize $[{\cal I}_{\rm gen}(t\to\infty)\neq0]$, since infinite-time averages \cite{Alessio2016, Mondaini2015, SM} are readily available from numerics.

Further characterization of the transition can be done by noticing that subdiffusive equilibration dynamics precedes the onset of the localized phase, which can be traced to the formation of bottlenecks of rare regions in the disordered potential, thereby hampering energy transport. These are often referred as Griffiths regions \cite{Griffiths1969,Potter2015,Vosk2015,Luitz2016}, and have been experimentally shown to lead to a power-law decay in time of the imbalance, ${\cal I_{\rm gen}}(t)\propto t^{-\xi}$, with a disorder-dependent exponent $\xi$, that vanishes at the transition $V_c$ \cite{Luschen2017,Kohlert2019}. Here, since we observe that the onset of localization depends on the energy of the system, the regime of manifestation of subdiffusive behavior is also $\varepsilon$-dependent. We further report in Fig.~\ref{fig:Fig3}A the disorder averaged imbalance at different times with energy resolution, using disorder strength $V/2\pi = 4, 16$ and $50\ {\rm MHz}$. Figure \ref{fig:Fig3}{B} displays similar results when focusing on the central part of the energy spectrum ($\varepsilon = 0.5$). A variety of MBL studies performed to date mostly focus on this general trend: the reduction of the decaying exponent $\xi$ under the increase of disorder. We extend this analysis to include energy dependence of this behavior. When checking the decay with time, subsequent to the initial dynamics displaying coherent oscillations (Figures \ref{fig:Fig2}{C} and \ref{fig:Fig3}{B}), we are able to extract the disorder controlled $\xi$ exponent, that is dependent on both energy density  and disorder (Fig.~\ref{fig:Fig3}{C}). The compilation of the values of $\xi$ (Fig.~\ref{fig:Fig3}C) shows that $V_c = V_c(\varepsilon)$: it is typically easier to trigger localization when the system is not at the regime of infinite temperatures.

The degree of control of our quantum processor enables direct determination of quantities intrinsically related to the time-evolved wavefunction, not previously investigated in experiments. As an example, we are able to trace the time evolution of the multiqubit probabilities related to each Fock state $|n\rangle$, $p_n(t) = |\langle n|\Psi_t\rangle|^2$, allowing the quantification of how fast $|\Psi_t\rangle$ spreads over the Hilbert space. A direct measure of this is provided by the participation ratio, ${\rm PR} = \left[\sum_n p_n^2(t)\right]^{-1}$, shown in Fig.~\ref{fig:Fig4}A, that compares $\varepsilon=0.15$ and $0.5$. The contrast is evident: The fast spread at the larger energy density is dictated by ergodic behavior, whereas at $\varepsilon=0.15$, the slower growth of PR suggests the onset of non-ergodicity, with an order of magnitude difference at the longest experimental time.

A distinctive aspect of the MBL phase is the characteristic slow growth of the entanglement entropy ($\propto \log t$), a result directly connected to the exponential localization of its emerging set of local integrals of motion, the $\ell$-bits \cite{Nandkishore2015,Vosk2015,Abanin2019}. This should be contrasted with the ballistic spreading of entanglement occurring in thermalizing systems, with diffusive energy transport \cite{Kim2013}.
It has been tested numerically \cite{Bardarson2012}, and experimentally measured with quantum state tomography (QST)~\cite{Xu2018} and indirectly probed by $n$-point correlations~\cite{Lukin2019,Rispoli2019}.
Since QST is extremely time consuming for a large number of qubits, here we implement the quantum Fisher information,
defined as ${\cal F}_{Q}(t) = \langle{\cal I}_{\rm gen}^2(t)\rangle - \langle{\cal I}_{\rm gen}(t)\rangle^2$, for a lower bound of the entanglement entropy.
After the initial transient, ${\cal F}_{Q}$ also grows logarithmically in time within a regime governed by the MBL phenomenon. Figure \ref{fig:Fig4}{B} compares the time evolution at intermediate disorder strengths ($V/2\pi=16\ {\rm MHz}$) with energy densities $\varepsilon =$~0.15 and 0.5. In the latter, the fast growth of ${\cal F}_{Q}(t)$ approaching the regime of saturation indicates thermal behavior, whereas at smaller $\varepsilon$'s, the growth is logarithmically slow, suggesting localization.

By using a highly programmable quantum processor with 19 superconducting qubits, we provide a phase diagram of the MBL transition with unprecedented level of detail, resolving the energy dependence of the critical disorder. This is done by checking the evolution of a generalized imbalance, that encodes information on the initial state preparations. Exact numerical characterization of the emulated system is in excellent agreement with the experimental realization, suggesting the coexistence of localization and thermalization for the same disorder amplitude, albeit at different energy densities. Furthermore, our processor is scalable, with the potential to probe the yet unsolved question of the possible existence of mobility edges. The present experimental results suggest so, but a proper scaling with larger number of qubits in the regime where numerical simulation using classical computers becomes intractable, could possibly settle down this question.

\section*{Acknowledgments}
\noindent RM thanks discussions with Rosario Fazio and Marcos Rigol. Devices were made at the Nanofabrication Facilities at
Institute of Physics in Beijing and National Center for Nanoscience
and Technology in Beijing.
{\bf Funding:} Supported by National Natural Science
Foundation of China (Grants No. NSAF-U1930402, 11674021, 11974039, 11851110757, 11725419 11434008, and 11904145), and National Basic
Research Program of China grants (Grants No. 2017YFA0304300 and 2016YFA0300600). {\bf Author contributions:} C.C. and R.M. proposed the idea; C.C. and Z.-H.S. performed the numerical simulation; Q.G. conducted the experiment; H.L. and D.Z. fabricated the device; R.M., C.C., Q.G., Z.-H.S., H.F., and H.W. cowrote the manuscript; and all authors contributed to the experimental setup, discussions of the results, and development of the manuscript.
{\bf Competing interests:} Authors declare no competing interests.
{\bf Data and materials availability:} All data needed to evaluate the
 conclusions in the paper are present in the paper or the
 supplementary materials.

\bibliography{MainText_bib}

\clearpage

\begin{figure*}[h!]
 \includegraphics[width=.8\textwidth]{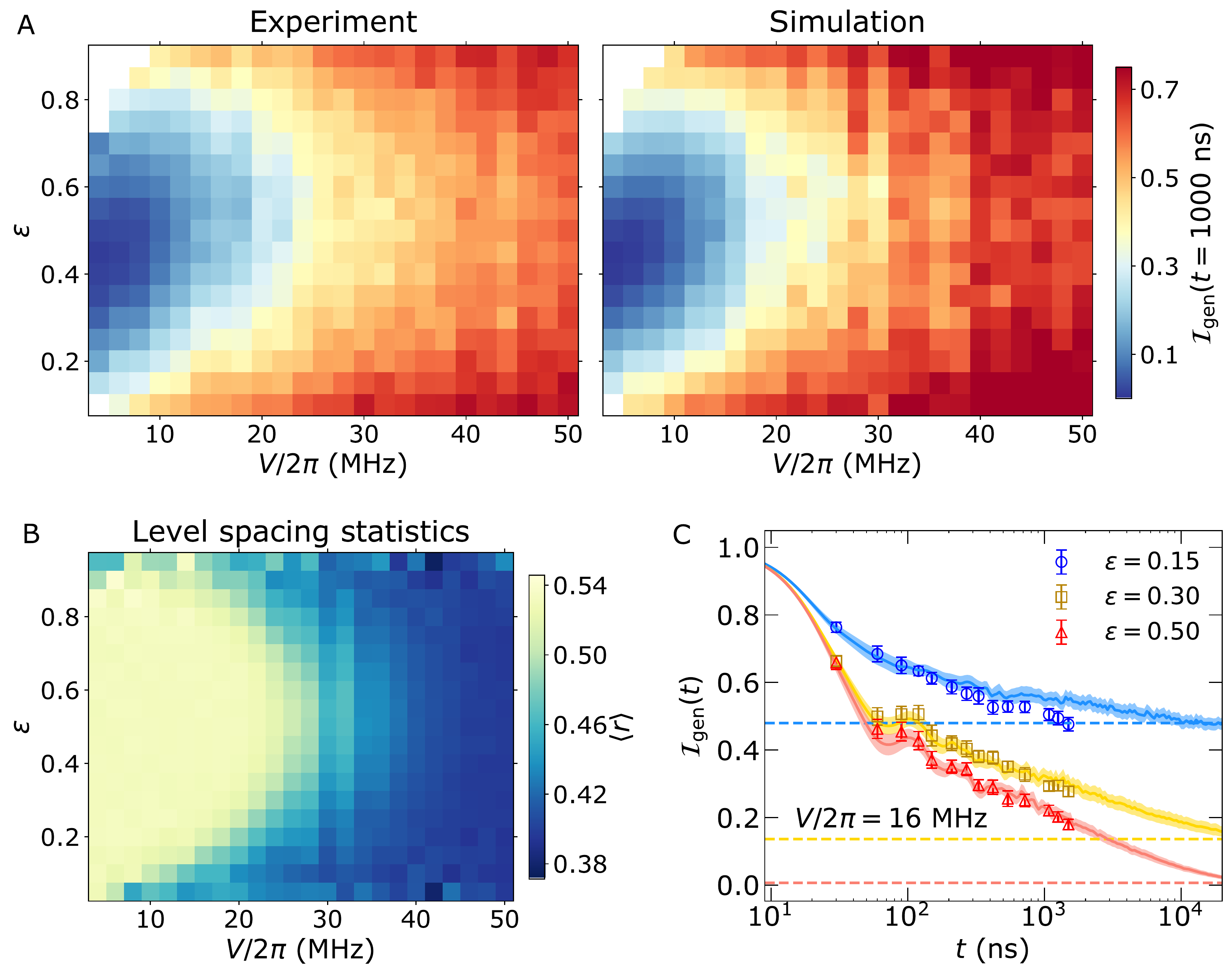}
 \vspace{-0.2cm}
 \caption{{\bf Many-body localization phase diagram -- Experiment vs. simulation.} ({\bf A}) Disorder averaged imbalance $\mathcal{I}_{\rm gen}$ measured at $t = 1000$~ns (left) in comparison with numerical simulation of the $\mathcal{I}_{\rm gen}(t)$ dynamics taking into account all device parameters except decoherence (right). Experimental data are obtained for $k = 20$ combinations of the random disorder realizations and the corresponding initial (Fock) states, encompassing a broad range of energies. ({\bf B}) Numerically extracted MBL phase diagram, illustrated by the average value of the ratio of adjacent gaps $\langle r\rangle$ as functions of both disorder $V$ and energy density $\varepsilon$. The calculation is based on the analysis of the energy level repulsion, which demonstrates striking similarity as the $\mathcal{I }_{\rm gen}$ data in (\textbf{A}) obtained via the dynamical process. Both experimental and numerical simulation data indicate the concomitant manifestation of extended and localized states at a given disorder strength $V$ ($10\lesssim V/2\pi\lesssim 30$~{MHz}) as allowed by the large Hilbert space of 19 qubits. ({\bf C}) Direct comparison between experiment (markers) and numerical simulation (lines) for $\mathcal{I}_{\rm gen}(t)$ at selected energy density $\varepsilon$ values as indicated, where $V/2\pi = 16$~MHz. For comparison, the infinite time averages (see Supplementary Material for the calculation) are shown as horizontal dashed lines. Error bars (experiment) and shadings surrounding the lines (simulation) denote the standard error of the statistical mean.}
 \label{fig:Fig2}
\end{figure*}

\begin{figure*}[h!]
 \includegraphics[width=0.95\textwidth]{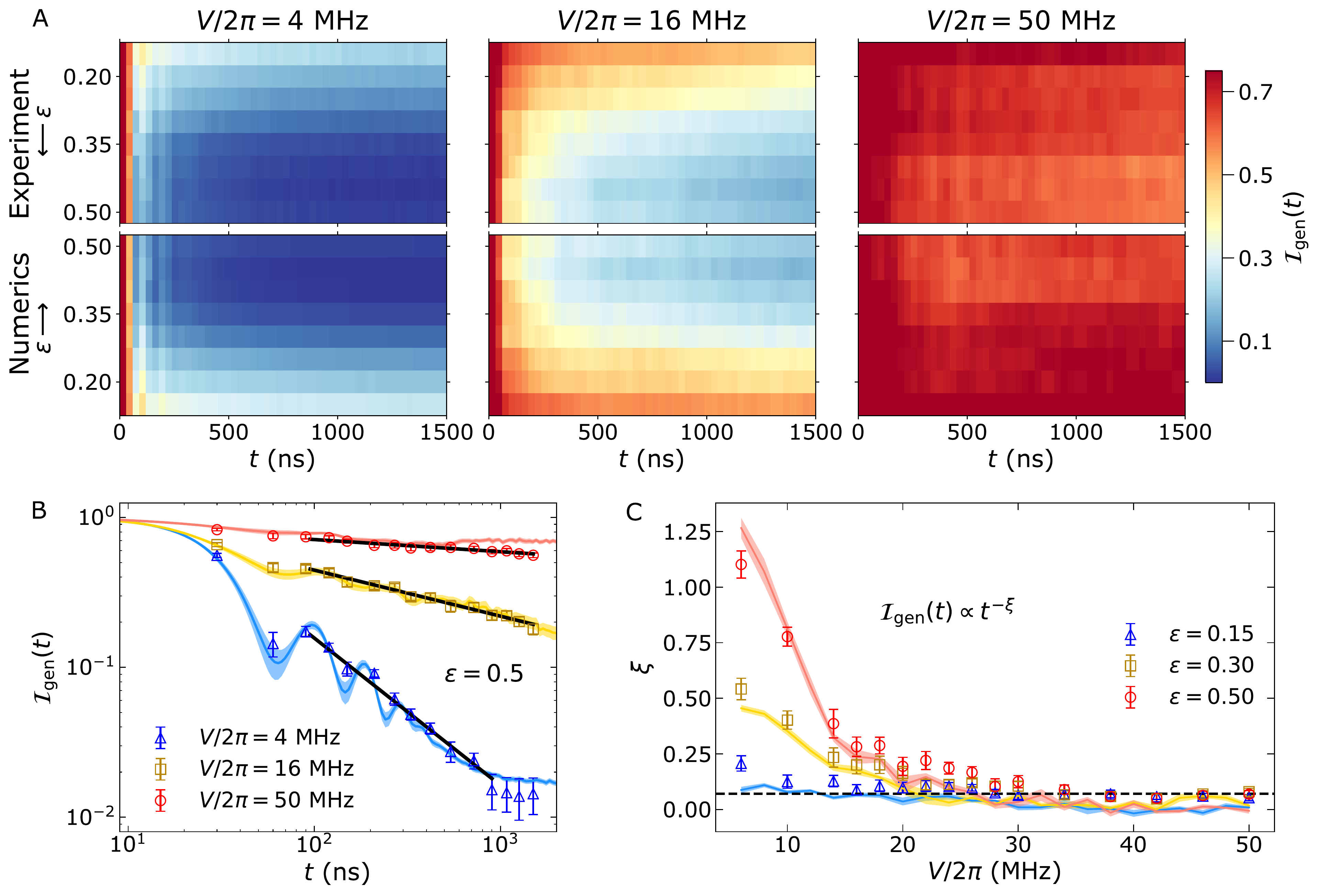}
 \vspace{-0.2cm}
 \caption{{\bf Real-time dynamics of imbalance -- Experiment vs. simulation.} ({\bf A}) Upper panels: Time-dependence of the disorder averaged imbalance $\mathcal{I}_{\rm gen}$ measured up to $t = 1500$~ns, for quenches at different energy densities $\varepsilon$ and disorder amplitudes, $V/2\pi$ = 4, 16, and 50 MHz. Lower panels: Numerical simulation of the $\mathcal{I}_{\rm gen}$ dynamics taking into account all device parameters except decoherence in comparison with experiment. Here data for half of the spectrum, which centers at the energy density $\varepsilon$ = 0.5, are plotted for visual contrast. ({\bf B}) Dynamics of $\mathcal{I}_{\rm gen}$ and its power-law decay in time, i.e., $\mathcal{I}_{\rm gen}(t)\propto t^{-\xi}$, at different disorder amplitudes $V$ as indicated. As $V$ increases, the exponent $\xi$ reduces towards zero and the system approaches the MBL phase featuring subdiffusive energy transport. Markers (Shaded lines) are experimental (simulation) data, and straight lines are fits to the experimental data in the time range 100 to $\sim$1000(1500)~ns for $V/2\pi=4$~(16,50)~MHz. Error bars and shaded regions represent the standard error of the mean. ({\bf C}) Exponent $\xi$ as a function of $V$ obtained from fitting the $\mathcal{I}_{\rm gen}(t)$ data as exemplified in (\textbf{B}). The critical disorder $V_c$, where $\xi$ approaches a finite but small constant as guided by the horizontal dashed line (estimated as the average value of $\xi$ at all experimentally chosen energy densities for the largest disorder amplitudes considered $V/2\pi \in$ [38~MHz, 50~MHz]), is associated to the onset of localization and also suggests $\varepsilon$-dependence. Error bars denote the uncertainty of the fit and lines interpolate the fitted exponents from numerical simulation. At large disorder amplitudes $V$, the small but nonzero values of $\xi$ by numerical simulation are due to finite size of the system. \\
 }
 \label{fig:Fig3}
\end{figure*}

\begin{figure*}[h!]
 \includegraphics[width=.7\textwidth]{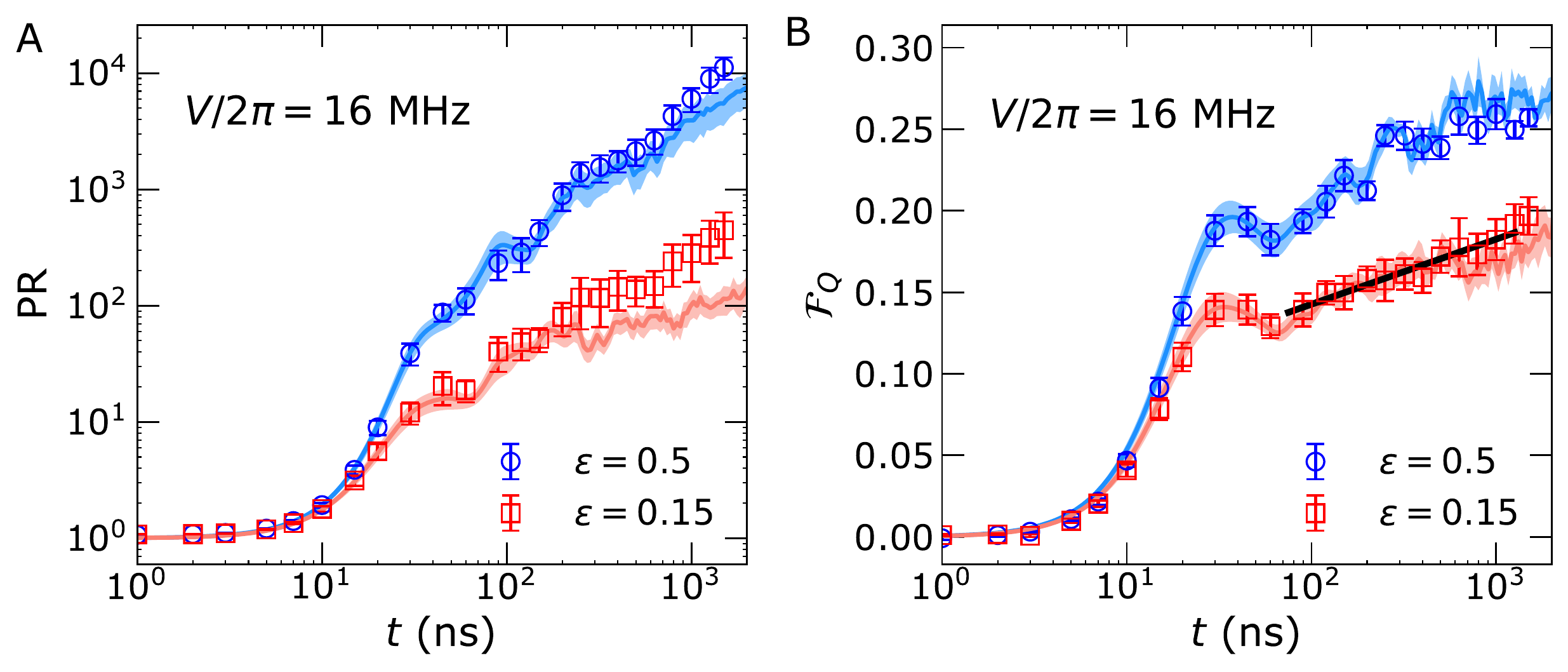}
 \vspace{-0.2cm}
 \caption{{\bf Wavefunction-related measurement and minimal entanglement quantification.} ({\bf A}) Average participation ratio (PR) in the Fock basis as a function of the evolution time $t$, at different energy densities $\varepsilon$ as indicated. Both experimental (markers) and numerical simulation (shaded lines) data are obtained for $k = 10$ combinations of the random disorder realizations and the corresponding initial (Fock) states whose ${\rm PR}(t=0)\simeq1$. The growth of PR at $\varepsilon=0.5$, the center of the spectrum (Fig.~\ref{fig:Fig2}A), is much faster than those at other energy densities away from the center. ({\bf B}) Time-dependence of the experimental quantum Fisher information ${\cal F}_{Q}$ (markers) in comparison with numerical simulation (shaded lines) at different energy densities as indicated. The $\mathcal{F}_{Q}(t)$ data, processed from the same multiqubit probabilities for plots in ({\bf A}), demonstrate a characteristic logarithmic growth in time after an initial transient, which signifies the MBL phase at small energy densities, e.g., $\varepsilon = 0.15$. In contrast, the $\mathcal{F}_{Q}$ data show a faster increase approaching saturation in a much shorter time scale at energy densities close to the center of the spectrum, e.g., $\varepsilon = 0.5$. In both panels $V/2\pi = 16$~MHz and error bars (shading regions) represent the standard error of the mean. Here we post-select the measured multiqubit probabilities for conservation of total excitations before calculating the PR and $\mathcal{F}_Q$ values.
 }
 \label{fig:Fig4}
\end{figure*}

\end{document}


\title{Supplementary Material for \\
Observation of energy resolved many-body localization}

\author{Qiujiang Guo}
\thanks{Those authors contributed equally to this work.}
\affiliation{Interdisciplinary Centre for Quantum Information and Zhejiang Province Key
	Laboratory of Quantum Technology and Device, Department of Physics, Zhejiang University,
	Hangzhou 310027, China}

\author{Chen Cheng}
\thanks{Those authors contributed equally to this work.}
\affiliation{School of Physical Science and Technology, Lanzhou University, Lanzhou 730000, China}
\affiliation{Beijing Computational Science Research Center, Beijing 100094, China}

\author{Zheng-Hang Sun}
\thanks{Those authors contributed equally to this work.}
\affiliation{Institute of Physics, Chinese Academy of Sciences, Beijing 100190, China}
\affiliation{CAS Center for Excellence in Topological Quantum Computation, School of Physical Sciences, UCAS, Beijing 100190, China}

\author{Zixuan Song}
\author{Hekang Li}
\author{Zhen Wang}
\author{Wenhui Ren}
\author{Hang Dong}
\affiliation{Interdisciplinary Centre for Quantum Information and Zhejiang Province Key
	Laboratory of Quantum Technology and Device, Department of Physics, Zhejiang University,
	Hangzhou 310027, China}

\author{Dongning Zheng}
\affiliation{Institute of Physics, Chinese Academy of Sciences, Beijing 100190, China}
\affiliation{CAS Center for Excellence in Topological Quantum Computation, School of Physical Sciences, UCAS, Beijing 100190, China}

\author{Yu-Ran Zhang}
\affiliation{Theoretical Quantum Physics Laboratory, RIKEN Cluster for Pioneering Research, Wako-shi, Saitama 351-0198, Japan}

\author{Rubem Mondaini}
\email{rmondaini@csrc.ac.cn}
\affiliation{Beijing Computational Science Research Center, Beijing 100094, China}

\author{Heng Fan}
\email{hfan@iphy.ac.cn}
\affiliation{Institute of Physics, Chinese Academy of Sciences, Beijing 100190, China}
\affiliation{CAS Center for Excellence in Topological Quantum Computation, School of Physical Sciences, UCAS, Beijing 100190, China}

\author{H. Wang}
\email{hhwang@zju.edu.cn}
\affiliation{Interdisciplinary Centre for Quantum Information and Zhejiang Province Key
Laboratory of Quantum Technology and Device, Department of Physics, Zhejiang University,
Hangzhou 310027, China}

\maketitle
\section{Device information}
As shown in Fig.~\textcolor[rgb]{1,0,0}{2A} of the main text, our device features 20 transmon qubits interconnected by a central bus resonator $\mathcal R$. Each qubit, labeled as $Q_j$ with $j\in \{1,2,\cdots,20\}$, is a frequency-tunable nonlinear LC resonator consisting of a flux-biased Superconducting Quantum Interference Device (SQUID) and a shunted capacitance. Twenty control lines are used to dynamically tune qubit frequencies (Z lines) and sixteen lines are used to drive the $|0\rangle \leftrightarrow |1\rangle$ transitions of the qubits (XY lines). Two readout lines are used to detect the joint states of the qubits, with each one shared by ten readout resonators ($R_j$), and each $R_j$ is dispersively coupled to $Q_j$. $\mathcal R$ is a  half-wavelength superconducting coplanar waveguide resonator, with a fixed resonant frequency of $\omega_R/2\pi\approx 5.248$ GHz. $\mathcal R$ has 20 side arms, and each arm capacitively couples to a qubit. Therefore, arbitrary two or more qubits on the device can interact with each other as mediated by $\mathcal R$, yielding the all-to-all connectivity which is ideal for simulating quantum many-body physics.

The key point of an analog quantum simulator lies in constructing its effective Hamiltonian that can be mapped onto the emulated system~\cite{RevModPhys.86.153}. For the specific topic that is being pursued in this experiment, we design our device so that the nearest neighbor coupling terms dominate over the long range interactions in the effective Hamiltonian, which is highly different from the 20-qubit device used in Ref.~\cite{Song2019}. As shown in the device image, we have shortened the separations between the neighboring qubits to enhance the direct couplings.

In this experiment we use 19 out of the 20 qubits. The unused qubit, labeled as $Q_{20}$, is unbiased and sitting close to its sweet point. Qubit performance is summarized in Table \textcolor[rgb]{1,0,0}{S1}. Detailed information of wirings and calibration routines can be found in Ref.~\cite{Song2018,Song2019,Xu2018}.

\begin{table}[htbp]
    \setlength{\columnsep}{8pt}
	\centering
	{\tabcolsep 0.13in  \begin{tabular}{cccccccccc}
    	\hline
    	\hline
	     &$\omega_{j, {\rm idle}}/2\pi$&$T_{1j, \rm idle}$&$\overline{T}_{1j, \rm operation}$&$T_{2j,\rm idle}^{*}$&$g_j/2\pi$&$\omega_{r,j}/2\pi$&$F_{0,j}$&$F_{1,j}$\\
	        &(GHz)&($\mu$s)&($\mu$s)&($\mu$s)&(MHz)&(GHz)& \\
	\hline
	$Q_1$   &4.450&$\sim$41&$\sim$31&$\sim$1.4&18.0&6.549&0.968&0.910\\
	$Q_2$   &4.069&$\sim$50&$\sim$50&$\sim$1.8&18.0&6.677&0.959&0.910\\
	$Q_3$   &4.612&$\sim$46&$\sim$46&$\sim$2.0&18.3&6.795&0.961&0.918\\
	$Q_4$   &4.109&$\sim$61&$\sim$59&$\sim$2.0&17.2&6.737&0.937&0.914\\
	$Q_5$   &4.220&$\sim$55&$\sim$51&$\sim$1.9&16.6&6.607&0.970&0.933\\
	$Q_6$   &5.064&$\sim$40&$\sim$55&$\sim$2.4&16.0&6.581&0.986&0.928\\
	$Q_7$   &4.676&$\sim$62&$\sim$55&$\sim$2.1&16.4&6.696&0.954&0.915\\
	$Q_8$   &5.095&$\sim$28&$\sim$68&$\sim$2.6&17.8&6.753&0.980&0.937\\
	$Q_9$   &4.163&$\sim$37&$\sim$41&$\sim$1.7&17.5&6.642&0.958&0.928\\
	$Q_{10}$&4.567&$\sim$30&$\sim$58&$\sim$2.0&17.5&6.522&0.970&0.898\\
	$Q_{11}$&4.951&$\sim$19&$\sim$37&$\sim$1.7&20.3&6.529&0.970&0.882\\
	$Q_{12}$&4.663&$\sim$38&$\sim$48&$\sim$1.1&17.4&6.656&0.966&0.926\\
	$Q_{13}$&4.990&$\sim$46&$\sim$60&$\sim$2.1&18.2&6.771&0.962&0.910\\
	$Q_{14}$&5.161&$\sim$23&$\sim$54&$\sim$1.6&16.7&6.715&0.988&0.923\\
	$Q_{15}$&5.013&$\sim$44&$\sim$55&$\sim$2.8&16.1&6.588&0.972&0.935\\
	$Q_{16}$&4.639&$\sim$35&$\sim$30&$\sim$2.0&17.4&6.679&0.979&0.931\\
	$Q_{17}$&5.045&$\sim$21&$\sim$54&$\sim$2.9&18.4&6.742&0.986&0.888\\
	$Q_{18}$&4.240&$\sim$38&$\sim$30&$\sim$1.0&18.1&6.619&0.963&0.917\\
	$Q_{19}$&4.585&$\sim$40&$\sim$37&$\sim$1.8&18.1&6.504&0.938&0.904\\
	$Q_{20}$&6.063&$\sim$36&       -&$\sim$4.4&16.1&6.563&0.978&0.860\\

		\hline
		\hline
	\end{tabular}}
	\caption{\label{Table S1}\textbf{Qubit Performance}. $\omega_{j, {\rm idle}}$ is the typical value of the idle frequency for $Q_j$, where we prepare initial states for different energy densities. $T_{1j, \rm idle}$ and $T_{2j,\rm idle}^{*}$ are $Q_j$'s energy relaxation time and Ramsey (Gaussian) dephasing time, respectively. $\overline{T}_{1j, \rm operation}$ is the average energy relaxation time over a spectrum range from 4.63 GHz to 4.73 GHz, which covers the frequencies where different disordered potentials $V_m$ are applied during the multiqubit interacting dynamics. $g_j$ is the coupling strength between $Q_j$ and $\mathcal R$. $\omega_{r,j}$ is approximately the resonant frequency of $Q_j$'s readout resonator $R_j$. $F_{0, j}$~($F_{1, j}$) is the typical measured probability when $Q_j$ is prepared in state $|0\rangle$ ($|1\rangle$), which is used to correct the raw data for elimination of the readout errors~\cite{Xu2018}.}
\end{table}

\section{Effective Hamiltonian}
In the rotating-wave approximation, with the direct coupling $\lambda_{mn}$ between $Q_m$ and $Q_n$ included, the system Hamiltonian for 19 qubits is given by
$$\frac{H_0}{\hbar}=\omega_Ra^{\dagger}a+ \sum_{m=1}^{19}[\omega_m\sigma^{+}_{m}\sigma^{-}_{m}+g_m(\sigma^ {+}_{m}a^{-}+ \sigma^{-}_{m}a^{+})]+\sum_{m=1}^{18}\sum_{n=m+1}^{19}\lambda_{mn}(\sigma^{+}_{m} \sigma ^{-}_{n}+\sigma^{-}_{m}\sigma^{+}_{n}),$$
where $\omega_m$ is $Q_m$'s frequency, $a^{\dagger}$ $(a)$ is the creation (annihilation) operator of bus resonator $\mathcal{R}$, $\sigma_m^{+}$ $(\sigma_m^{-})$ is the raising (lowering) operator for qubit $Q_m$, and $g_m$ is the coupling strength between $Q_m$ and $\mathcal{R}$. When all the qubits are far detuned from $\mathcal{R}$ by the same amount of $\Delta = \omega_m-\omega_R$, which satisfies the relation $|\Delta|\gg |g_j|$, any pair of qubit $Q_m$ and $Q_n$ are coupled through virtual photon exchange mediated by $\mathcal{R}$ with an effective coupling strength $J_{mn}^{SE}=g_m g_n/\Delta$~\cite{zheng2000}. If $\mathcal{R}$ is initialized in ground state, the Hamiltonian is transformed to
$$\frac{H_0^\prime}{\hbar}=\sum_{m=1}^{18}\sum_{n=m+1}^{19}(\lambda_{mn} + J^{SE}_{mn})(\sigma^{+}_{m} \sigma ^{-}_{n}+\sigma^{-}_{m}\sigma^{+}_{n})+\sum_{m=1}^{19}h_m\sigma^{+}_{m}\sigma^{-}_{m},$$
where $h_m=g_m^2/\Delta$ is vacuum-induced frequency shift which can be eliminated by slightly adjusting qubit frequencies. Thus, with a random disordered local potential $V_m$ introduced to each qubit $Q_m$, the time evolution for the whole system is governed by the effective Hamiltonian
$$\frac{H}{\hbar}=\sum_{m=1}^{18}\sum_{n=m+1}^{19} J_{mn}(\sigma^{+}_{m} \sigma ^{-}_{n}+\sigma^{-}_{m}\sigma^{+}_{n})+\sum_{m=1}^{19}V_m\sigma^{+}_{m}\sigma^{-}_{m}.$$
While the effective coupling strength $J_{mn}$ = $\lambda_{mn} + J_{mn}^{SE}$ remains almost constant once the detuning $\Delta$ is fixed, the random disordered local potential $V_m$ is highly programmable by adjusting the qubit frequencies.

Typically, the direct coupling $\lambda_{mn}>0$ whose value decays quickly as the on-chip distance between $Q_m$ and $Q_n$ increases, and the resonator mediated interaction $J_{mn}^{SE} < 0$ at the experimentally selected $\Delta/2\pi\approx-568$~MHz. For neighboring pairs, direct coupling $\lambda_{mn}$ plays a leading role with a positive amplitude of several megahertz; for pairs beyond nearest neighbor, $J_{mn}^{SE}~(\sim -0.5$~MHz) dominates in magnitude. The value of $J_{mn}$ can be inferred by systematically measuring the $Q_m$-$Q_n$ on-resonance swap dynamics as a function of $\Delta$ centered around $\Delta/2\pi\approx-568$~MHz,
shown in Fig.~\textcolor[rgb]{1,0,0}{2B} of the main text. Therefore, together with a given disorder realization $\{{V_m}\}$, all the parameters of the effective Hamiltonian $H$ are settled down, which allow us to calculate the corresponding extremal eigenvalues, $E_{\rm min}$ and $E_{\rm max}$ for our protocol.

\section{Post-selecting the measurement results}

\begin{figure*}[t]
  \centering
  \includegraphics[width=6.5in,clip=True]{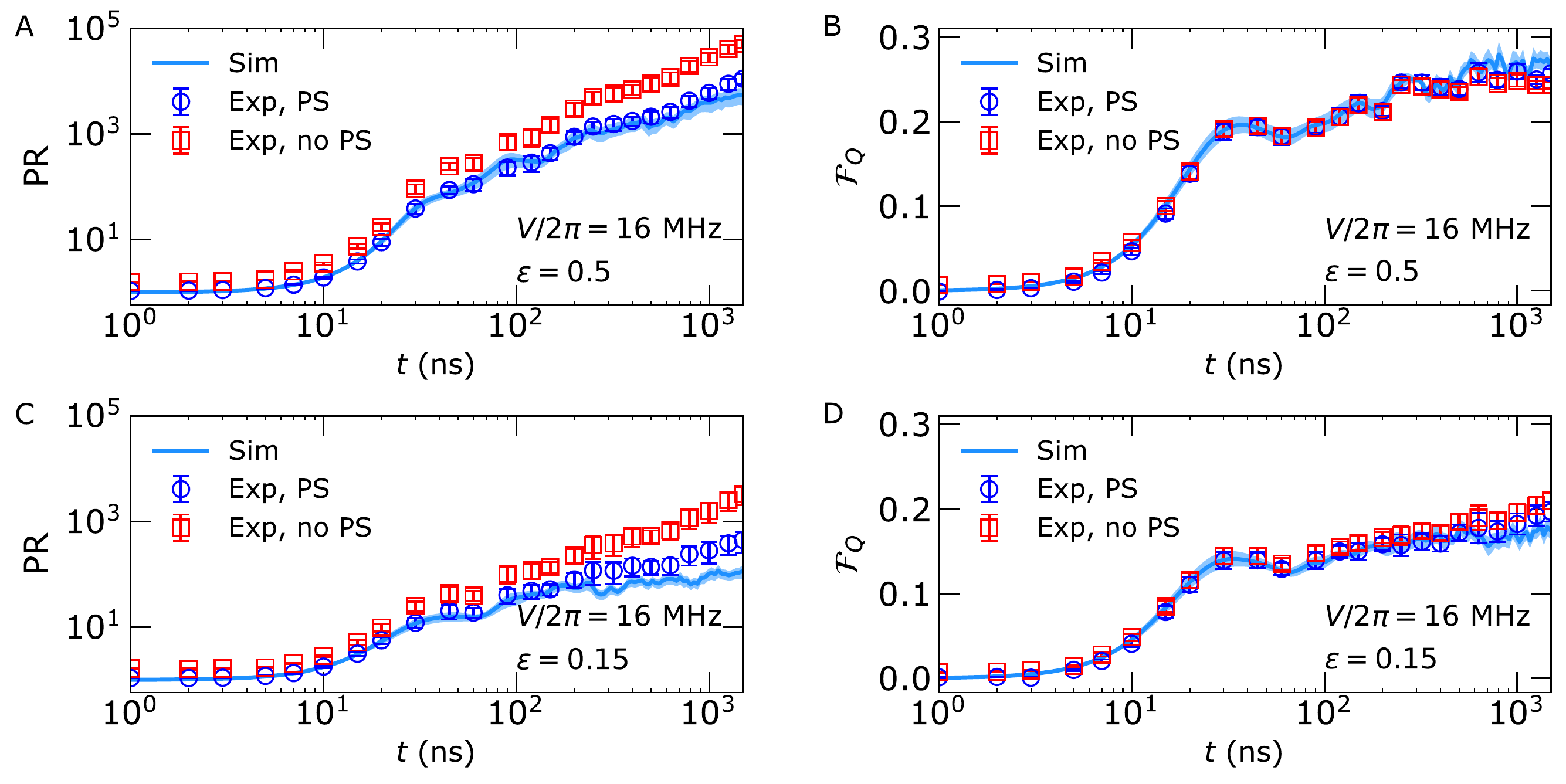}\caption{\label{figS2}{\bf Effect of post-selection.} Plotted are the experimental data for PR (left) and $\mathcal{F}_Q(t)$ (right) with (blue circles) and without (red squares) post-selection (PS), in comparison with the numerical results (blue lines). The effect of energy relaxation is eliminated by the post-selection process, but dephasing remains which could slightly affect the experimental data~\cite{Xu2018}.}
\end{figure*}

The Hamiltonian in Eq.~\textcolor[rgb]{1,0,0}{1} of the main text is an XY model, which conserves the total photon number (9 photons) of the initial state during the time evolution. Although the experimental evolution time up to 1500 ns is far less than the qubit energy relaxation times listed in Table \textcolor[rgb]{1,0,0}{S1}, with 19 qubits the leakage out of the 9-photon subspace may not be negligible, which could cause an overestimate of participation ratio (PR). Thus, we post-select the measured probabilities within the 9-photon subspace for the data shown in Fig.~\textcolor[rgb]{1,0,0}{5} of the main text. The comparison between the experimental results with and without post-selection at $\epsilon=0.15$ and 0.5 are shown in Fig.~\textcolor[rgb]{1,0,0}{\ref{figS2}}. We observe a much-improved agreement between the experimental and the numerical results with post-selection. We note that the experimental value of quantum Fisher information, $\mathcal{F}_Q(t)$, is not that sensitive to the post-selection process.

\section{Procedure of selecting initial states}
\begin{figure}[h!]
 \includegraphics[width=0.9\columnwidth]{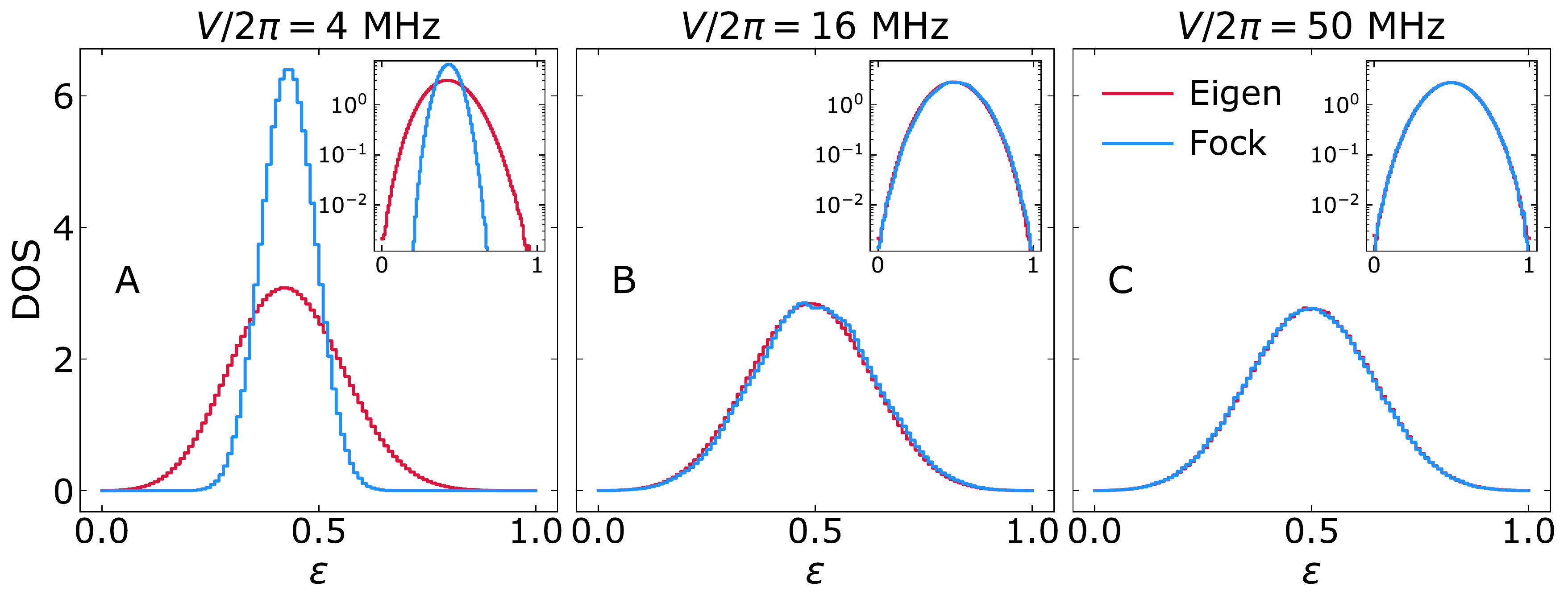}
 \vspace{-0.3cm}
 \caption{{\bf Comparison of density of states -- usefulness of our protocol.} The averaged density of states for both the Fock and eigenstates of the Hamiltonian, averaged over 20 and 10 disorder realizations, respectively. (A), (B) and (C) denote the increasing disorder amplitudes ($V/2\pi = 4, 16$ and $50\ {\rm MHz}$). Insets present the corresponding data in log scale to facilitate visualization.}
\label{SM_Fig_DOS}
\end{figure}

Our protocol to investigate the energy resolved many-body localization is based on the selection of initial product (Fock) states and posterior checking of the resulting generalized imbalance. As we indicate in the main text, the initial state encodes the energy of the (isolated) system in the quench protocol, and the usefulness of our approach is clear if the span in energy of the Fock states is compatible with the one of the actual Hamiltonian spectrum. To test this, we plot the density of states (DOS) for both the space of all possible Fock states as well as for the eigenstates of the Hamiltonian in Fig.~\textcolor[rgb]{1,0,0}{\ref{SM_Fig_DOS}}, for different disorder amplitudes. At large and intermediate disorder strengths ($V/2\pi = 50$ and $16\ {\rm MHz}$, respectively), the two DOS coincide almost perfectly, which attests the efficiency of our approach in investigating the different possible energies and its localization properties. Only at fairly small disorder amplitudes (see $V/2\pi = 4\ {\rm MHz}$ data) there is a small discrepancy, with a very small occurrence of product states corresponding to the far extremes of the spectrum.
As a result, the number of unitary evolution realizations for specific energy density and disorder strength is not necessarily the same as the number of disordered samples, especially at the edges of the spectrum for very small disorders. This is the reason we have blank regions in the color diagrams of the imbalance at small values of $V/2\pi$.

\section{Generalized imbalance at infinite times}

A possible critique one may incur on whether the experimentally measured (generalized) imbalance is significative of the onset of many-body localization, is that, while in principle infinite time evolution is required to demonstrate the breakdown of thermalization, in the current experiment we are only able to tackle finite times. Therefore, it is helpful to estimate how representative are finite time results in approaching $t\to\infty$. The latter can be evaluated by the following steps, in case the system's Hamiltonian is amenable to be numerically diagonalized.

For generic quantum systems, it is possible to write down (in units where $\hbar=1$) the time evolution of a pure state, quenched by the Hamiltonian $H$ one is emulating, in terms of its eigenstates as:
\begin{equation}
|\psi(t)\rangle = \sum_{\alpha=1}^{\cal D} e^{-{\rm i} E_\alpha t} c_\alpha |\alpha\rangle,
\end{equation}
where ${\cal D}$ is the Hilbert space dimension, and $c_\alpha \equiv \langle\alpha|\psi(0)\rangle$, is the overlap of each eigenstate ($H|\alpha\rangle = E_\alpha |\alpha\rangle$) and the initially prepared $|\psi(0)\rangle$. The time-dependence of an observable, can be thus written in terms of its matrix elements in the eigenbasis of $H$,
\begin{equation}
\langle {\cal O}(t)\rangle = \sum_{\alpha,\beta=1}^{\cal D} e^{{\rm i}(E_\alpha-E_\beta)t}c_\alpha^* c_\beta {\cal O}_{\alpha\beta},
\end{equation}
with ${\cal O}_{\alpha\beta}=\langle\alpha|{\cal O}|\beta\rangle$. Under very general conditions as, e.g., the energy spectrum not possessing non-coincidental degeneracies, and with off-diagonal matrix elements of the observable (${\cal O}_{\alpha\beta \ (\alpha\neq\beta)}$) being exponentially small in comparison to diagonal ones
(${\cal O}_{\alpha\alpha}$), it is possible to show that the average value at infinite-times, $\langle {\cal O}(t\to\infty)\rangle = \lim_{t\to\infty} \frac{1}{t} \int_0^t \langle {\cal O}(t^\prime)\rangle{\rm d}t^\prime$, converges to the ``diagonal ensemble'' \cite{Rigol2008}:
\begin{equation}
\langle {\cal O}_{\rm DE}\rangle = \sum_{\alpha=1}^{\cal D} |c_\alpha|^2 {\cal O}_{\alpha\alpha}.
\end{equation}

These values were used in Fig.~\textcolor[rgb]{1,0,0}{3C} in the main text as horizontal dashed lines, and help in identifying the saturation of both numerical and experimental results of the generalized imbalance. Following this expression, we further include below in Fig.~\textcolor[rgb]{1,0,0}{\ref{fig_SM:IG_DE}}, the energy-density resolved phase diagram of the generalized imbalance averaged at infinite times, which can be directly contrasted with the right panel of Fig.~\textcolor{red}{3A} in the main text, originally shown at the same time scale of the experiment, $t=1000\ {\rm ns}$. The generic `\textit{D}'-shape structure is still qualitatively similar, thus ruling out the possibility of being a mere finite-time effect.

\begin{figure}[h!]
  \centering
 \includegraphics[width=0.6\columnwidth]{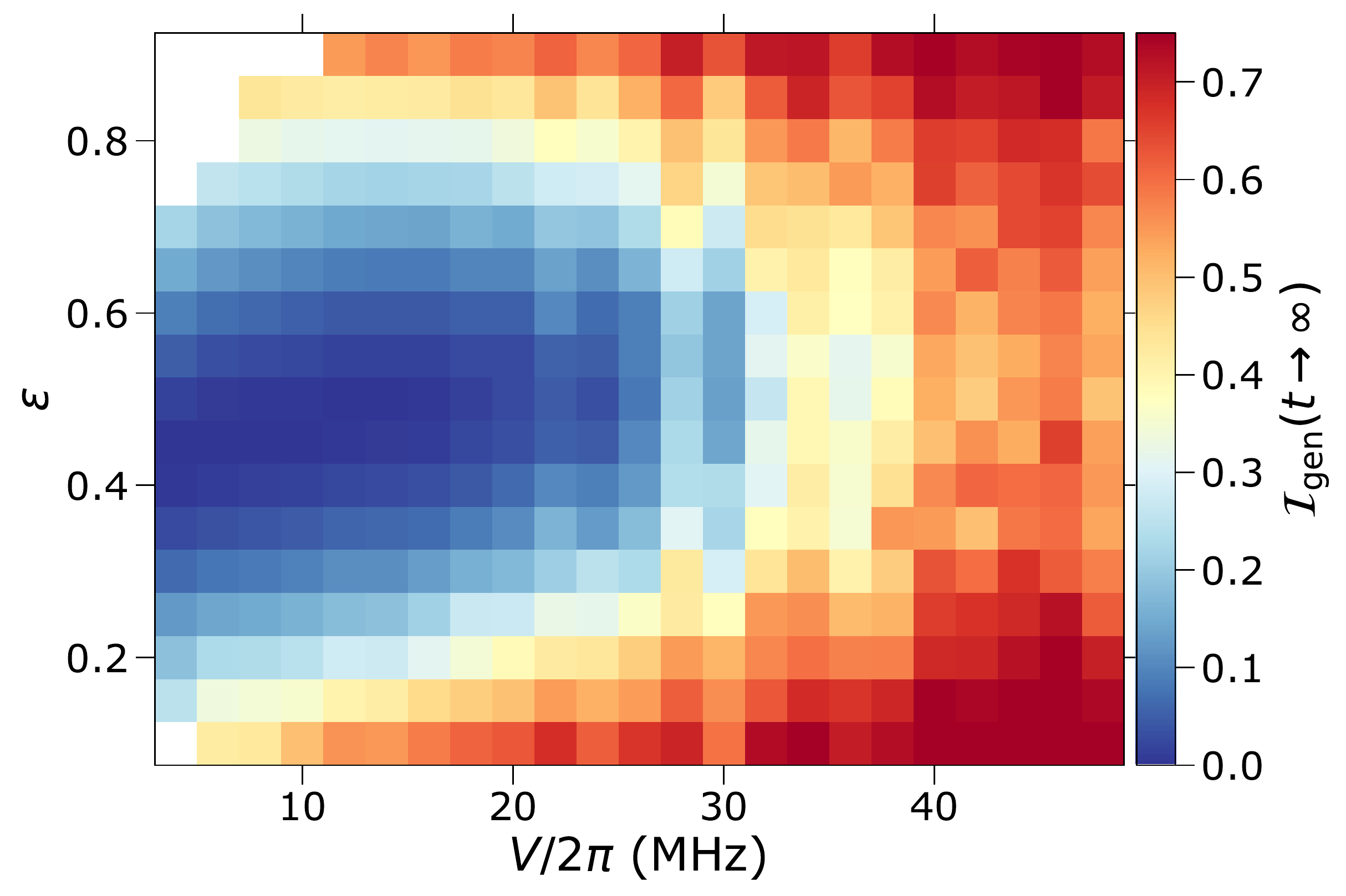}
 \vspace{-0.3cm}
 \caption{{\bf Generalized imbalance at infinite times.} The energy density -- disorder amplitude diagram of the diagonal ensemble (see text) for the generalized imbalance ${\cal I}_{gen}$. This corresponds to the infinite time average for this quantity, and shows that the inhomogeneous behavior at different energies is still preserved at this limit $t\to\infty$.}
\label{fig_SM:IG_DE}
\end{figure}

\section{Finite size analysis}
The existence of many-body mobility edges (MBMEs) is an important, but controversial issue in the MBL community. While it is generally found in finite systems, some believe that if one is able to approach the thermodynamic limit, the thermalizing effects that the ``bath'' formed by extended states becomes more effective, and would ultimately rule out the appearance of the mobility edge. However, since numerical studies are restricted to fairly small system sizes, its existence in the thermodynamic limit can be hardly proved/disproved using classical computers.

On the other hand, in a quantum ``computer'' or other kind of quantum devices, the size of the many-body system that can be emulated is not restricted by the exponentially large Hilbert space. Thus, quantum emulation dealing with a larger number of qubits has a chance in providing a final answer to the debate on MBME. In fact, the qubit number in our current experiment does not surpass the present numerical limits. Nevertheless, we can use our current platform to preliminary point out the route to be taken when dealing with quantum processors with a larger number of qubits, and already indicate the trend within our present experiment.

A simple finite-size analysis can be performed by removing the coupling of some of the qubits with the remaining ones, effectively reducing the total number of qubits in the emulation of the unitary evolution. One of the drawbacks of this approach is that the number of couplings in the Hamiltonian (Eq. \textcolor[rgb]{1,0,0}{1}) decreases at a step corresponding to the number of remaining qubits, due to the presence of the all-to-all couplings in our device. However, since the long-range couplings are an order of magnitude smaller than the nearest-neighbor ones, this does not significantly compromise a preliminary scaling study.

\begin{figure}[h!t]
 \includegraphics[width=0.85\columnwidth]{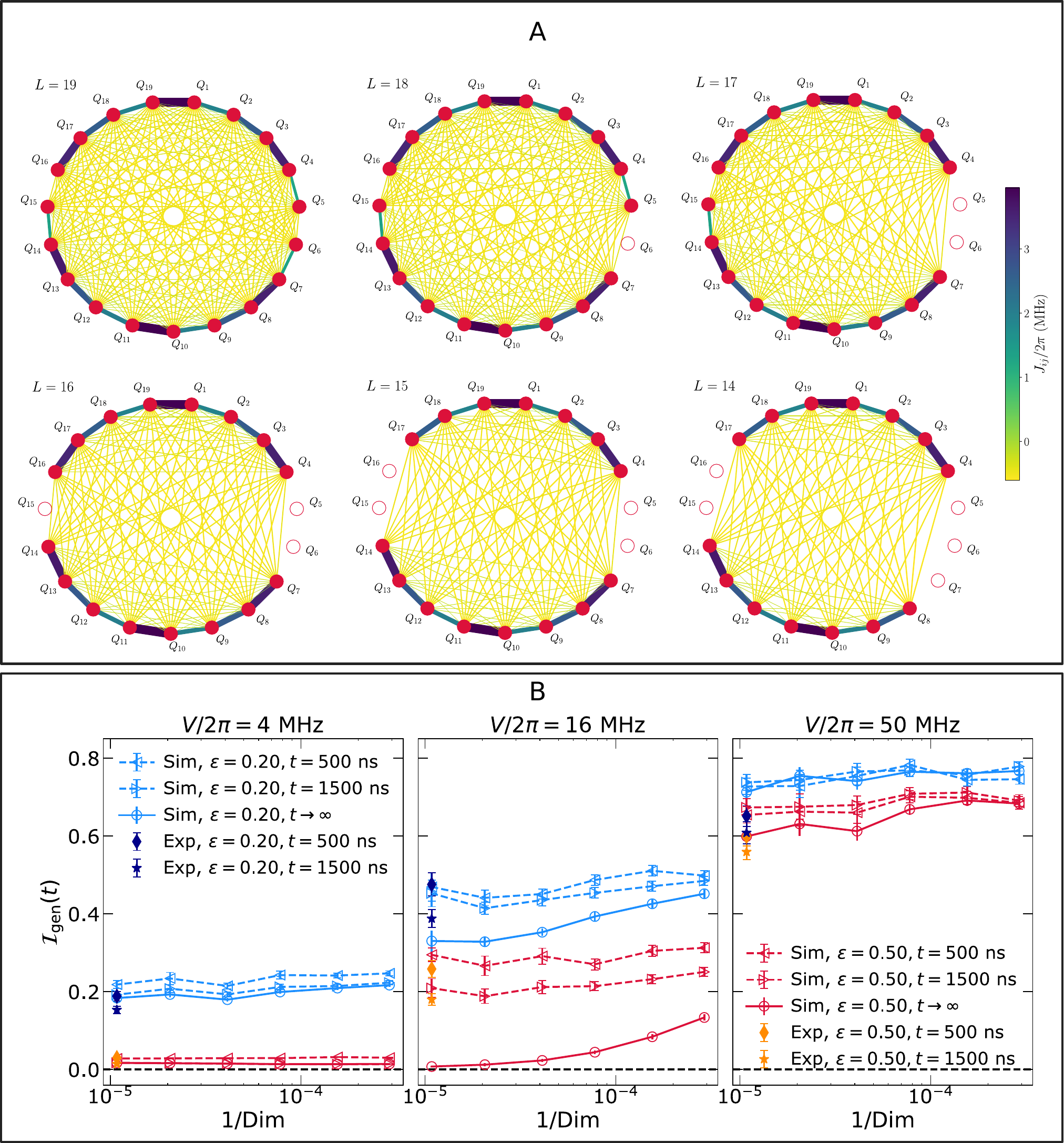}
 \vspace{-0.2cm}
 \caption{{\bf Finite size analysis: Imbalance $\cal{I}_\textrm{gen}$ with different numbers of qubits.} (A) Cartoons schematically displaying the qubits and their corresponding couplings (color and line thickness associated with its amplitude) for decreasing numbers of qubits, from 19 (originally used in the experiment) to 14 qubits. (B) The generalized imbalance ${\cal I}_{\rm gen}$ dependence on the Hilbert space dimension for small, intermediate and large disorder amplitudes, $V/2\pi=4$, $16$ and $50\ {\rm MHz}$, respectively. The numerical (experimental) results are denoted by empty (full) markers. The numerically obtained infinite time averages (diagonal ensemble) are connected by solid lines. The contrast in $\cal{I}_\textrm{gen}$ between the two energy densities $\varepsilon = 0.2$ and $0.5$ (empty markers connected by blue and red solid lines, respectively) is clear at the intermediate disorder strength, $V/2\pi=16\ {\rm MHz}$, with a seemingly finite and vanishing imbalance, respectively, up to the largest Hilbert space dimension we can reach, when approaching the $t\to\infty$ limit.}
\label{SM_Fig_finitesize}
\end{figure}

This removal of qubits is exemplified in Fig.~\textcolor{red}{\ref{SM_Fig_finitesize}A}, starting from the original 19 qubits device, scaling down to 14. In terms of the Hilbert space dimension, it amounts for a drop of the range from 92,378 down to 3,432 states, if preserving the total spin in $z$-direction $S^{tot}_z=0\,(-1)$ for an even (odd) total number of qubits. Following this, we report in Fig.~\textcolor{red}{\ref{SM_Fig_finitesize}B}, the generalized imbalance for increasing Hilbert space dimensions, focusing on small  ($V/2\pi=4\ {\rm MHz}$, left panel), intermediate  ($V/2\pi=16\ {\rm MHz}$, central panel) and large  ($V/2\pi=50\ {\rm MHz}$, right panel)
disorder amplitudes. These are the same amplitudes highlighted in Fig.~\textcolor{red}{4A} in the main text, when focusing on the time dependence of the imbalance. 
Here we preferentially use numerical simulation when dealing with smaller numbers of qubits and can draw the following conclusions.

\noindent\textbf{(i)} At strong disorder, $V/2\pi=50\ {\rm MHz}$, both energy densities $\varepsilon = 0.2$ and $0.5$ result in a converged finite imbalance, signifying that the MBL phase permeates the whole energy spectrum. This is different from the cases under small and intermediate $V$'s (see below), where the contrast in $\cal{I}_\textrm{gen}$ between $\varepsilon = 0.2$ and $0.5$ become more clear.

\noindent\textbf{(ii)} For $V/2\pi = 4\ {\rm MHz}$, the imbalance possess a small dependence on the Hilbert space size, acquiring a finite value when approaching the thermodynamic limit for $\varepsilon = 0.2$, that is suggestive of the MBL manifestation. In the middle part of the spectrum, $\varepsilon = 0.5$, the imbalance is extremely small, indicating thermalization and loss of information of initial preparations. In both cases, the equilibration times are short, with the time scales $t=1500\ {\rm ns}$ already matching the infinite time average value (diagonal ensemble).

\noindent\textbf{(iii)} When $V/2\pi=16\ {\rm MHz}$, on the other hand, the equilibration times are much longer but the distinction between the two energy densities are also very clear. Although the imbalance at the time scales reachable by the experiment is finite for $\varepsilon = 0.5$, the corresponding average approaching $t\to\infty$
vanishes already for the current device. This is not the case for smaller energy densities such as $\varepsilon = 0.2$, where the system conserves a large imbalance at this limit. This preliminary contrast already highlights the possible manifestation of a mobility edge, and the scale of our platform for a regime with much larger number of qubits (where simulations using classical computers become unfeasible), after following this same analysis, may settle the current question on its possible existence.

\bibliography{SupMat_bib}